# Blueshift and Phase tunability in planar metamaterials with toroidal contribution

**Maria V. Cojocari[1,2], Kristina Schegoleva[1,2], Alexey A. Basharin[1,2,*]**

[1] *National University of Science and Technology (MISiS), The Laboratory of Superconducting metamaterials, 119049 Moscow, Russia*
[2] *National University of Science and Technology (MISiS), Department of Theoretical Physics and Quantum Technologies, 119049 Moscow, Russia*
**Corresponding author: alexey.basharin@gmail.com*

**We propose a model of tunable THz metamaterials. The main advantage is the blueshift of resonance and phase tunability due to toroidal excitation in planar metallic metamolecules with incorporated silicon inductive inclusions.**

Metamaterials are artificial structures with properties unattainable in natural media. Their exotic response is a promising platform for filling the THz frequency gap [1-3]. A separate class of metamaterials is the one with the toroidal response [4-15]. The toroidal observation is mediated by the excitation of currents flowing in inclusions of toroidal metamolecules, and resembles the poloidal currents along the meridians of gedanken torus [4,5]. Meanwhile, the destructive interference between the toroidal and electric dipole moments leads to lack the far- fields, but the fields in the metamolecule origin describing by δ-function [6,7]. Such fields configuration, referred as the anapole, allows to observe the new effect of Electromagnetically Induced Transparency (EIT) [6,7], provides an extremely high *Q*-factor in metamaterials [8], enables a cloaking for nanoparticles [9,10], and is a platform for confirmation of the dynamic Aharonov-Bohm effect [6,7].
Recently, it was demonstrated the anapole excitation in planar metamaterials, which enabled an extremely high *Q*-factor in microwave [8], which gives promising opportunities for tunable metamaterials due to the strong electromagnetic fields localization within metamolecules. In this paper, we consider a design of a metamaterial, discussed in Ref. 8,-in tunable regime. For this purpose, we incorporate the photoconductive silicon into the metamolecule (Fig. 1a) and study the response of the metamaterial in the THz regime. The silicon is simulated with the permittivity $\varepsilon_{Si}$=11.7 and a pump–power-dependent conductivity $\sigma_{Si}$, varied from $10^{-1}$ up to $10^6$ S/m, which means transition from dielectric to metallic state.
Metamolecules comprise of two split parts (Fig. 1a). The incident plane electromagnetic wave with electric field ***E*** aligned with the central wire excites conductive currents in each loop of the metamolecule. The currents form a closed vortex of magnetic field ***H***. As a result, such configuration of electromagnetic fields supports the toroidal dipole excitation, oscillating upward and downward within metamolecule. However, the electric dipole also arises in the metamolecule and maintains the anapole mode, in accordance with the destructive interference between electric and toroidal dipole moments. The advantage is a very narrow line in the transmission spectrum of a metamaterial [8]. Here, for the first time, we consider how the planar toroidal metamolecule can be exploited as a building block for terahertz modulators. We demonstrate the blueshift and the phase tunability regime and also discuss the role of losses.

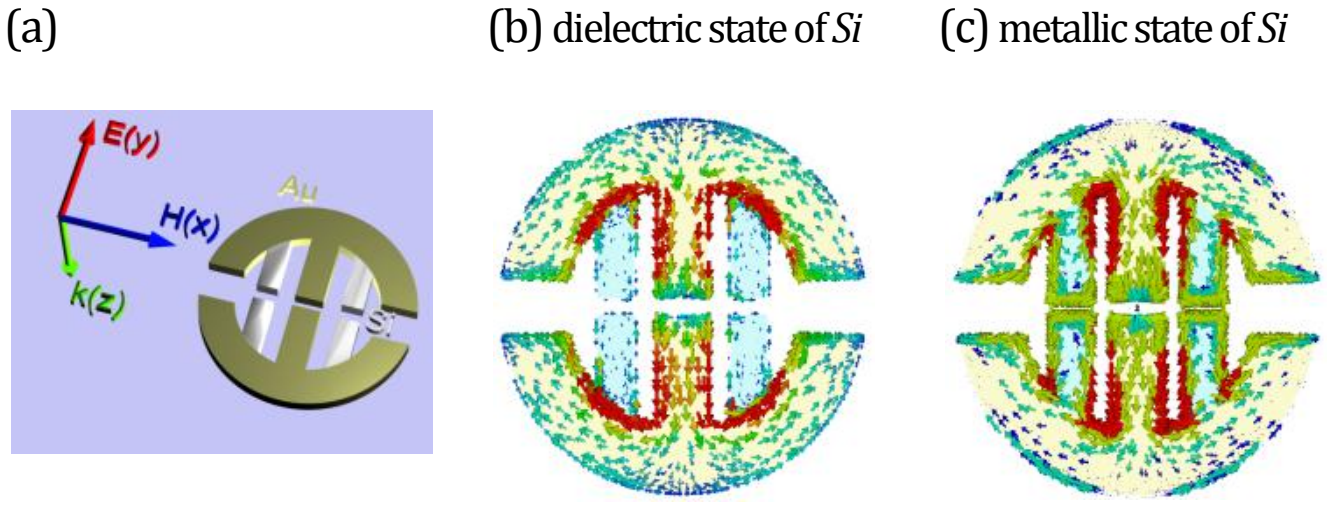

Fig.1. (a) Illustration of the proposed metamolecule, supporting toroidal dipolar excitation with incorporated silicon strips. The diameter of the metamolecule is 30 μm with central gap equals to 2 um and the lateral gaps 5 μm. The silicon strips have the gaps 1.4 um. (b)- conductive currents on the frequency 4.8 THz and for period *D*=37 μm, (c)- on the frequency 5.5 THz.

The metamaterials with incorporated silicon or gallium arsenide inclusions were discussed in details as elements of modulators. The phase tunability, blueshift, redshift as well as amplitude switching shown in THz frequency range [16-24]. Although the semiconductor inclusions play the role of capacitor elements or as the tunable metamaterial substrate.
Our advantage here is related with inductive tunability by changing the conductivity of silicon. This configuration introduces additional inductance in the metamolecule, and, due to increasing of silicon conductivity, we expect blueshift of resonance frequency by currents flowing along of silicon strips (Fig. 1c) instead of external metallic parts of metamolecule (Fig. 1b). Such incorporation of silicon decreases the electric dimensions of metamolecules and, thereby, blueshifts the resonance frequency.
The electromagnetic properties of metamaterial are computed by commercial Maxwell's equation solver HFSS. We suppose that all metamolecules periodically arranged in *x* and *y* directions with *D*=37 μm center- to- center separation. We study gold metamolecules with conductivity $\sigma_{gold}$=7x$10^6$ S/m. Fig. 2a shows the simulated transmission spectra for different conductivities of silicon. For small conductivities silicon is purely dielectric. Without pump beam the transmission minimum is characterized by resonance dip in vicinity of 4.56THz. With increasing conductivity the resonance is initially unchanged ($\sigma_{Si}$<$10^3$ S/m) (Fig. 3b). Since $\sigma_{Si}$=$10^3$ S/m we observe blueshift accompanied by increasing transmission intensity and attainable 5.25 THz on the $10^5$ S/m. Finally, for the metallic state of silicon $\sigma_{Si}$=$10^6$ S/m the transmission resonance achieves the maximum shift of 5.5 THz (Fig. 2a, 3b). However, we note the negligible small phase perturbation during $\sigma_{Si}$ variations (Fig. 2b).

We note that such blueshift is due to the contribution of silicon inclusions in overall distribution of currents. While at low conductivities ($\sigma_{Si}<10^3$ S/m) the currents flow along metallic parts of metamolecules (Fig. 1b), but with the transition to the metallic state ($\sigma_{Si}>10^3$ S/m) currents flowing along the silicon inclusions are dominate (Fig. 1c). This corresponds to the decreasing of electric size of metamolecule and to the blueshift of resonance frequency from 4.56 THz to 5.5 THz, that corresponds to 50% tunability (Fig. 2).

(a)

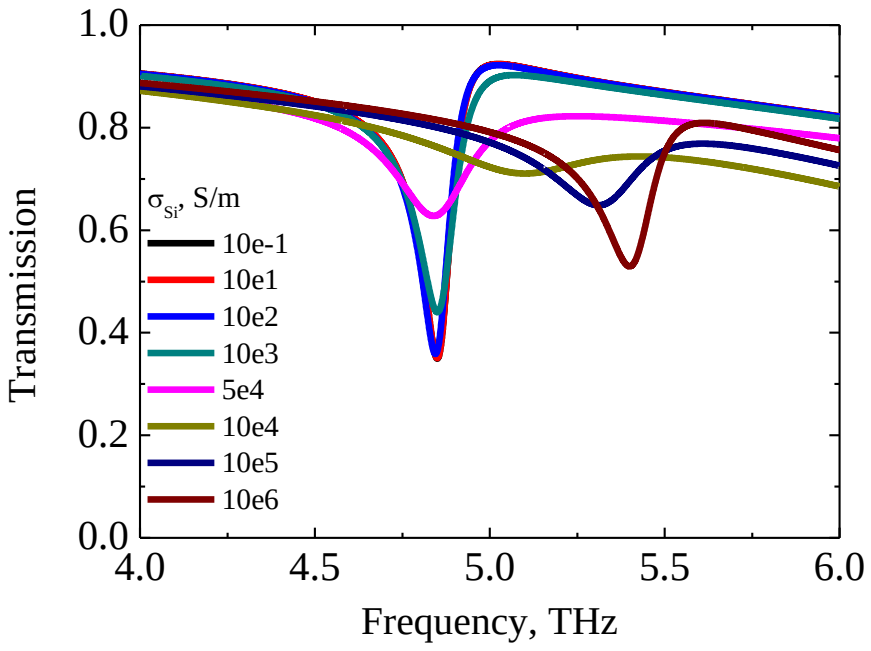

(b)

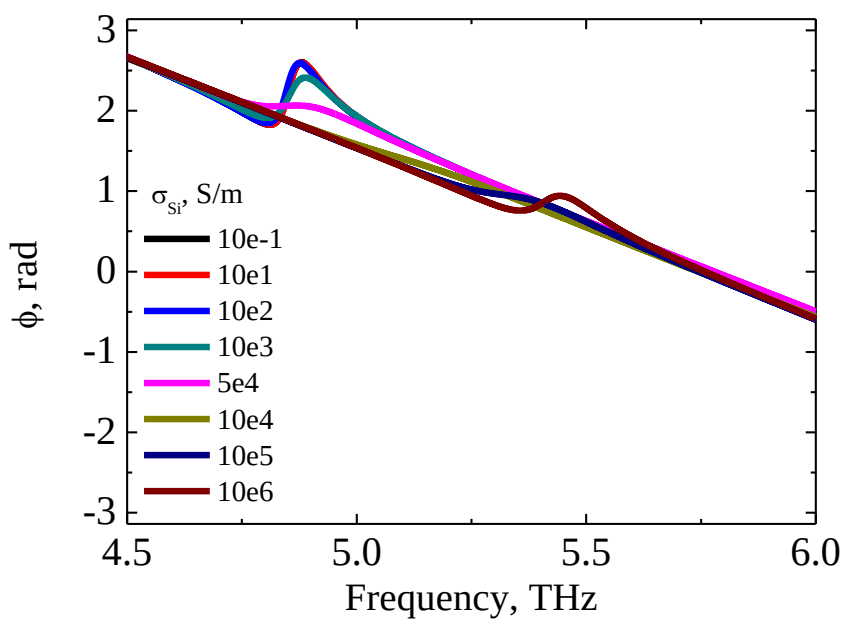

Fig. 2 shows the simulated transmission spectra for different conductivities of silicon dependence on frequency for the period of metamolecules $D$=37 μm. (a)- intensities and (b)- phases.

In order to evaluate the role of multipoles in resonances formation, we consider the energy of five strongest multipoles calculated from conductive and displacement currents extracted from the simulation (Fig. 3a). We plot normalized power radiated by electric **P**, magnetic **M**, toroidal **T**, electric quadropole **Qe** and magnetic quadropole **Qm** moments, corresponding to transmission dips depicted on Fig. 2a in dependence on silicon conductivities $\sigma_{Si}$.

For low pump illumination the silicon is purely dielectric and resonance mainly defined by electric and toroidal multipoles. However, with increasing of $\sigma_{Si}$ the decomposition between moments occurs and electric moment will dominate more strongly. Intensities of other multipoles are less than 100 times in comparison with electric and toroidal moments. Moreover, a significant difference between the electric and toroidal dipoles explains low Q-factor at frequencies ~5.5 THz and $\sigma_{Si}>10^4$ S/m compared with the higher Q-factor dip on the frequencies ~4.8 THz and $\sigma_{Si}<10^4$ S/m, for which the electric and toroidal dipoles have very close intensities to each other. Thus, we observe the blueshifted Fano-type resonance which is varied by transition of silicon conductivity from dielectric to metallic state.

(a)

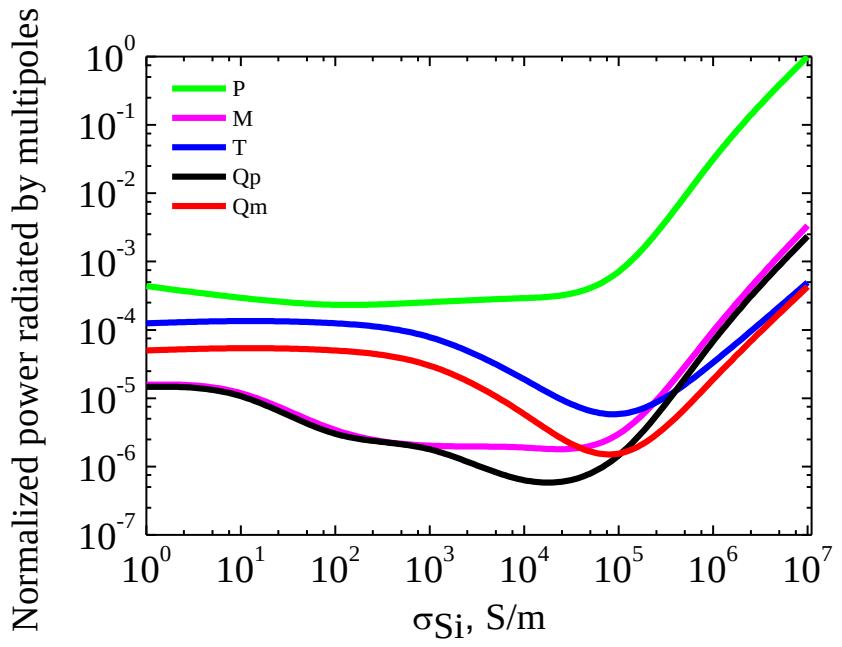

(b)

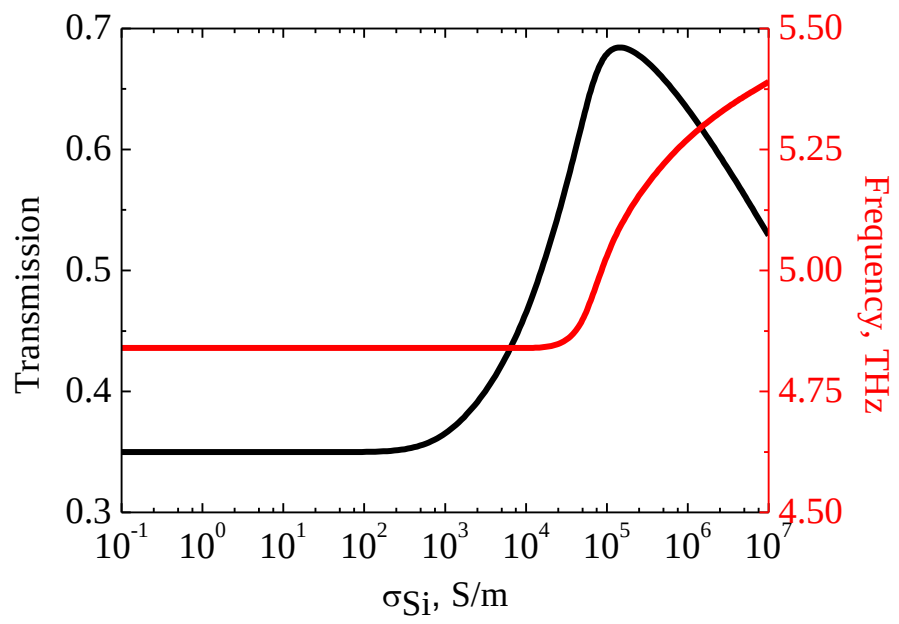

Fig. 3 shows (a)- contributions of the five strongest multipolar excitations to the electromagnetic response from metamaterial sample in *log* scale for the case of period $D$=37 μm and (b)- the transmission and resonance frequency depended on silicon conductivity $\sigma_{Si}$.

Interestingly, the proposed metamaterial can exhibit a phase tunability. For this purpose we assume distance between metamolecules centers $D$=30 μm. At low conductivities of silicon $\sigma_{Si}$ pronounced peak is significantly reduced to 0.5 (Fig. 4a). While with growing of conductivity $\sigma_{Si}$ one cannot be distinguished from the overall transmission curve close to the 6.5 THz. Thus, the amplitude tunability is impossible caused by resonance. However, we can make an assumption here. Dissipative losses in the metamolecules influence on the transmission amplitude, whereas large phase tunability is observed in order to transition from the dielectric state of silicon to the metallic state (See Summary). We achieve tunability ~2 rad, which is the benefit of proposed metamaterial as a platform for THz phase modulators (Fig. 5).

(a)

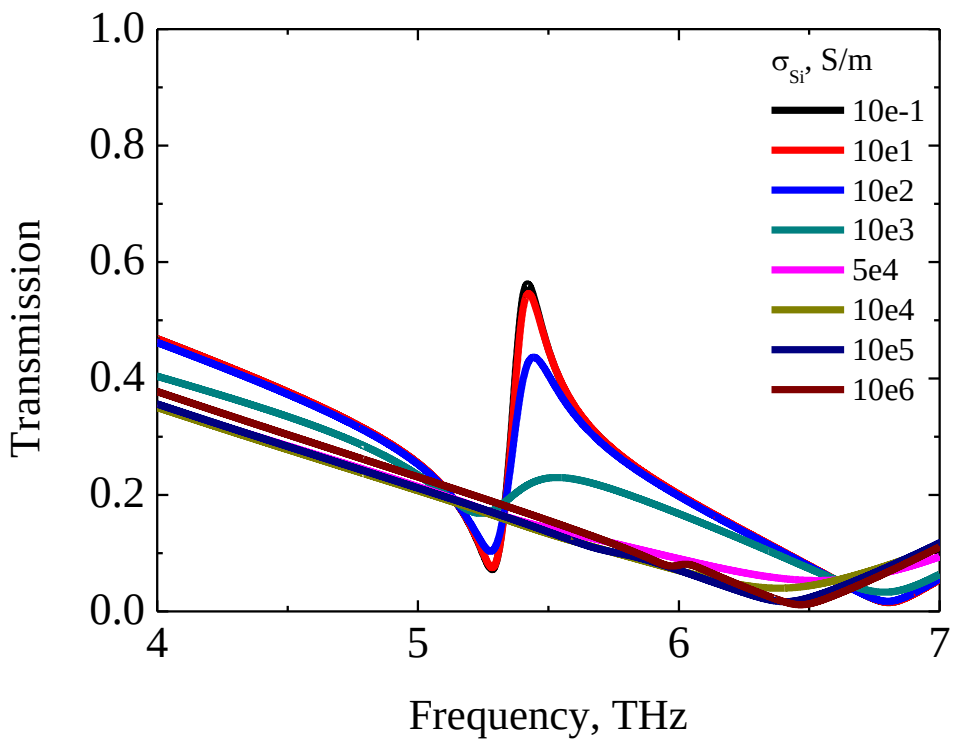

(b)

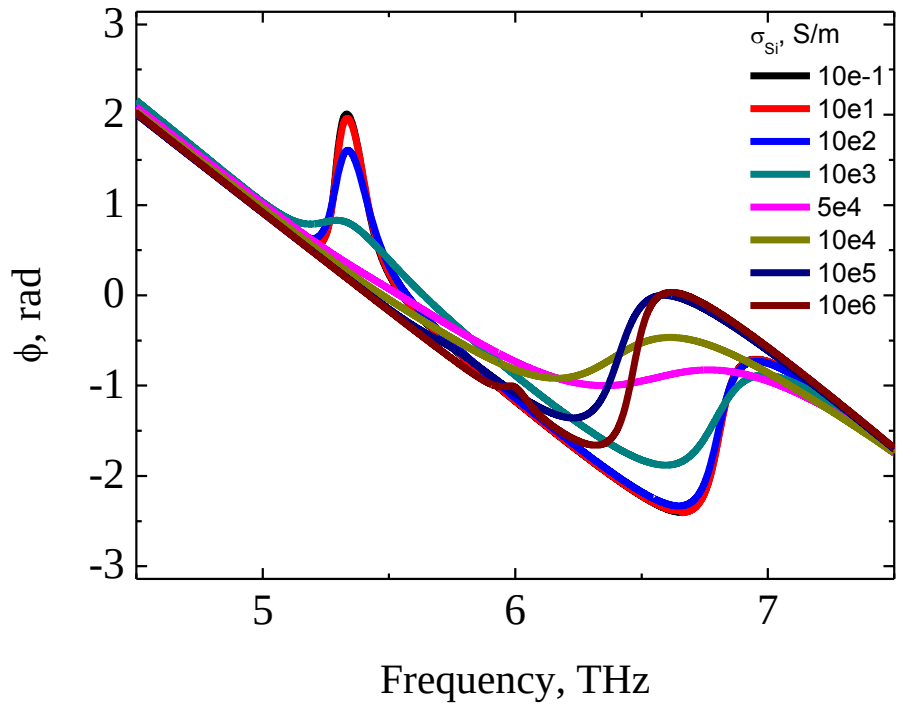

Fig. 4 shows the simulated transmission spectra for different conductivities of silicon dependence on frequency for the period of metamolecules $D$=30 μm. (a)- intensities and (b)- phases.

Indeed, the variation of silicon conductivity $\sigma_{Si}$ from 1 S/m to the $10^6$ S/m significantly changes the phase from -2 to 0 rad, associated with non-disturbing intensity in broad frequency range from 6.5- 7.2 THz (Fig. 4a,b). Moreover, on the Fig. 5 we generalize the tunability characteristics. We show the contribution of conductivity $\sigma_{Si}$ on the transmission phase and amplitude at 6.6 THz. For low conductivities ($<10^3$ *S/m*), the phase shift is almost constant. The significant phase shift occurs for $\sigma_{Si}>10^3$ *S/m*, which is linearly enhanced up to $10^6$ *S/m*, whereas the transmission amplitude is constant and very low during the $\sigma_{Si}$ variation.

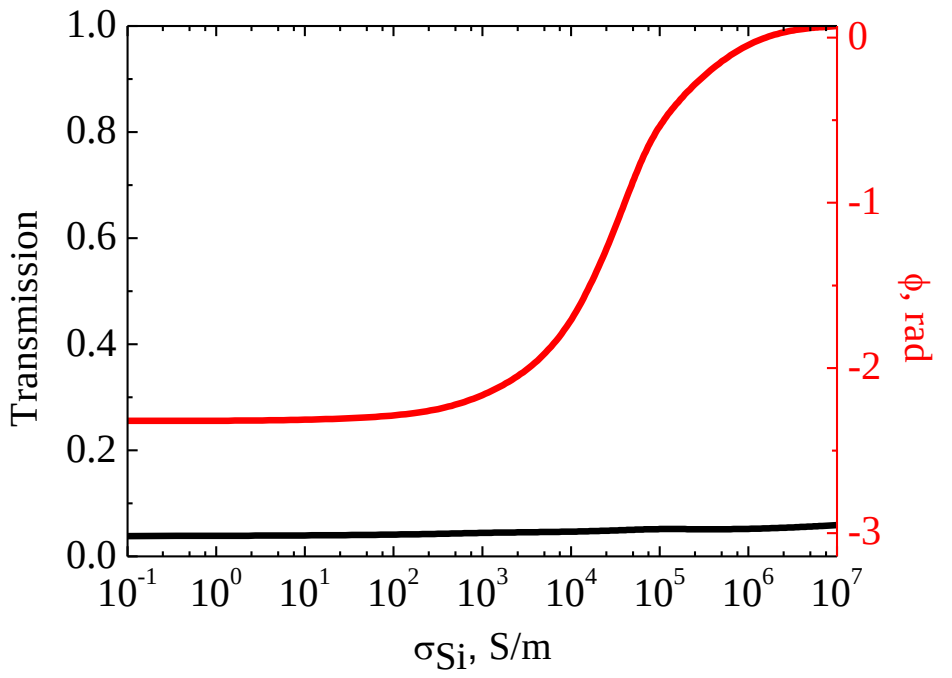

Fig. 5. The amplitude and phase tunability dependenced on silicon conductivity $\sigma_{Si}$ on the 6.5 THz for metamaterial with period $D$=30 μm.

Importantly, we also discuss the role of multipoles in forming the phase tunability. Indeed, the contributions of all multipoles are the same for varied conductivities $\sigma_{Si}$, the electric dipole moment dominates and toroidal intensities are very high in comparison with other multipoles at the frequency 6.6 THz (Fig. 6a), during the $\sigma_{Si}$, evaluation. Moreover, the phase behavior (Fig. 6b) of toroidal and electric dipoles reminds the graph of Fig. 5. Thus, the main origin of the phase tunability is the phase manner of multipoles instead of amplitude of them. Thus, the achieved phase tunability is very high ~2 rad. We expect that our metamaterial would operate with low pump intensity without high power sources, because we need act only on small silicon wires.

(a)

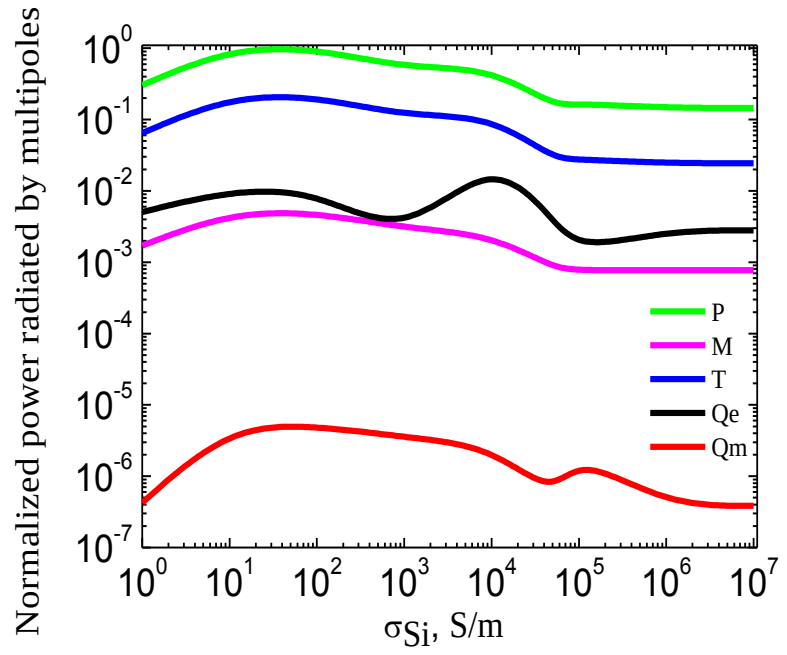

(b)

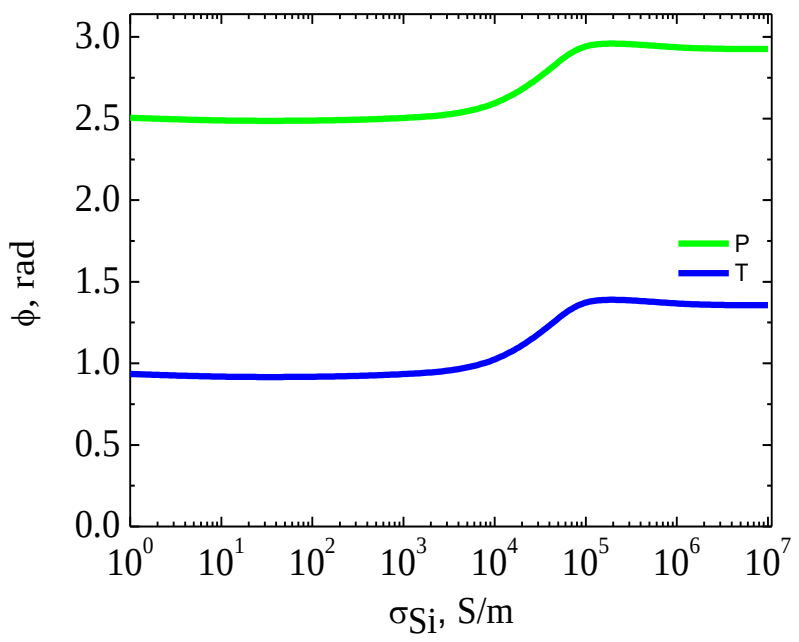

Fig. 6 (a) Contributions of the five strongest multipolar excitations to the electromagnetic response from metamaterial sample in *log* scale for the case of period $D$=30 μm. (b) The phases of the electric and the toroidal dipole moments vs frequency.

As a summary, we discuss the role of losses in proposed metamaterials. The actual losses of metals in THz frequency range can be crucial for discussed effects. Although metals in the microwave range can be considered as PEC, the real conductivities in the THz region limiting the $Q$-factor and destroy the high-$Q$ resonances. We represent on Fig. 7 the amplitude and phase tunability for the proposed metamaterial with $D$=30 μm. We plot graph for varied conductivity of metamolecules material $\sigma_{met}$, where $\Delta A=|A_{dielectic}-A_{metallic}|$ and $\Delta\phi=|\phi_{dielectic}-\phi_{metallic}|$, intensities and phases characterized the tunability or the difference between metallic and dielectric state of silicon. Decreasing of $\sigma_{met}$ reduces the resonance intensity, while the phase tunability is evident. It implies, that the dissipative losses destructively affect the transmission amplitude, while phase tunability is crucial for real conductivities of metals in THz regime ($\sigma_{met}<10^8$ S/m) and amplitude tunability is lack in this case. This explains why we observed significant phase tunability ~2 rad on Fig. 4b without amplitude tunability (fig. 4a, 5).

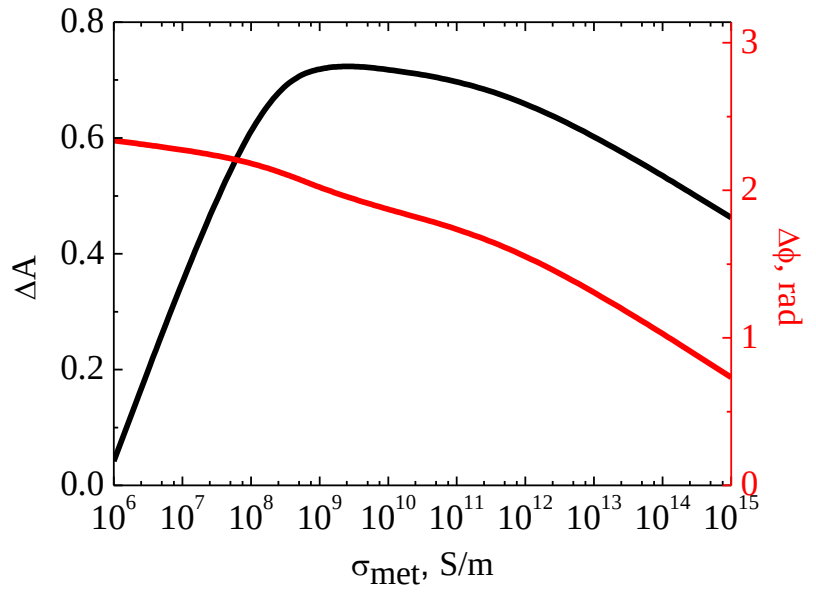

Fig. 7 shows the simulated tunability characteristics of intensity and phase difference of electromagnetic wave passed through metamaterial sample dependence on metamolecules conductivity $\sigma_{met}$ for period $D$=30 μm.

It is well known that the metamaterial response is limited by radiating and nonradiation losses. $Q$-factor of metamaterials is described as:

$Q=1/Q_{rad}+1/Q_{non}$, **(1)**

where $Q_{rad}$ is radiating losses and $Q_{non}$- is nonradiating or dissipative losses. While the radiating losses in toroidal metamaterials are low, we are limited mainly by Joule losses in ingredients, which can be compensated by exploiting superconducting or dielectric inclusions [25-27]. However, the formula (1) address that we have background for playing between $Q_{rad}$ and $Q_{non}$. Indeed, negligible mismatch between multipoles leads to the decreasing of $Q_{rad}$, but $Q_{non}$ has some frame for manipulation of losses in order to enhance the $Q$-factor. This fact explains why the transmission resonance for $D$=37 μm (Fig. 2a) is higher Q than for $D$=30 μm (Fig. 4a). In this way, we enhanced the radiation losses and as results, the interaction between electric, toroidal and other multipoles.

Recently, the design and fabrication of planar THz metamaterials and modulators based on them have been discussed intensively. Ref [16-24] suggested us the possibilities for successful implementation of proposed metamaterials in future.

In conclusion, we proposed a design of toroidal metamaterial base on planar inclusions with incorporated photoconductive silicon. Our finding shown deep blueshift and phase tunability realized due to changing of conductivity of silicon. We believe that our approach paves the way to realization of THz toroidal modulators.

**Funding.** NUST MISiS (K4-2015-031); RFBR (16-32-50139, 16-02-007)

**Acknowledgment**. We would like to thank Ilia Besedin and Igor Golovchanskiy for fruitful discussions. This work was supported by the Ministry for Education and Science of the Russian Federation, in the framework of the Increase Competitiveness Program of the National University of Science and Technology MISiS under contract numbers K4-2015-031, the Russian Foundation for Basic Research (Grant Agreements No. 16-32-50139 and No. 16-02-00789)